\documentclass[reprint,aps,prl,twocolumn,
superscriptaddress
]{revtex4}

\usepackage{graphicx}
\usepackage{color}
\usepackage{appendix}
\usepackage{times}
\usepackage{hyperref}

\newcommand{\id}[0]{\ensuremath{\mathbf{1}}}
\newcommand{\bra}[1]{\langle #1|}
\newcommand{\ket}[1]{|#1\rangle}

\newcommand{\abs}[1]{{\left\vert#1\right\vert}}

\newcommand{\kb}[2]{\ensuremath{\vert #1 \rangle \langle #2 \vert}}
\newcommand{\exx}[1]{\ensuremath{\langle #1 \rangle}}

\newcommand{\sz}[0]{\ensuremath{\mathbf{\sigma}_z}}

\newcommand{\an}[0]{\ensuremath{a}}
\newcommand{\cre}[0]{\ensuremath{a^\dagger}}
\newcommand{\no}[0]{\ensuremath{\cre \an}}

\newcommand{\ddt}[0]{\frac{\mathrm{d}}{\mathrm{d}t}}

\begin{document}

\title{Quantum thermometry using the ac Stark shift within the Rabi model}

\author{Kieran D. B. Higgins}
\email{kieran.higgins@materials.ox.ac.uk}
\affiliation{Department of Materials, Oxford University, Oxford OX1 3PH, United Kingdom}
\author{Brendon W. Lovett}
\affiliation{SUPA, Department of Physics, Heriot Watt University, Edinburgh EH14 4AS, United Kingdom}
\affiliation{Department of Materials, Oxford University, Oxford OX1 3PH, United Kingdom}
\author{Erik M. Gauger}
\email{erik.gauger@materials.ox.ac.uk}
\affiliation{Centre for Quantum Technologies, National University of Singapore, 3 Science Drive 2, Singapore 117543}
\affiliation{Department of Materials, Oxford University, Oxford OX1 3PH, United Kingdom}

\date{\today}

\begin{abstract}
A quantum two level system coupled to a harmonic oscillator represents a ubiquitous physical system. New experiments in circuit QED and nano-electromechanical systems (NEMS) achieve unprecedented coupling strength at large detuning between qubit and oscillator, thus requiring a theoretical treatment beyond the Jaynes Cummings model. Here we present a new method for describing the qubit dynamics in this regime, based on an oscillator correlation function expansion of a non-Markovian master equation in the polaron frame. Our technique yields a new numerical method as well as a succinct approximate expression for the qubit dynamics. We obtain a new expression for the ac Stark shift and show that this enables practical and precise qubit thermometry of an oscillator. 
\end{abstract}


\maketitle

The qubit-oscillator model has gone by many names in many fields, owing its tenacity to the breadth of its applicability: It is the simplest non-trivial model of the interaction between light and matter. At its inception it was used to describe the interaction of an atom with a magnetic field~~\cite{rabi36}, and referred to thereafter as the Rabi model. In the subsequent decades it has been extensively studied in quantum optics \cite{shore1993} and cavity QED \cite{raimond01}.  Physical chemists have used a `vibration-dimer' model to study the spectra of molecules \cite{fulton61}.  Applying the rotating wave approximation (RWA) to the Rabi model yields the Jaynes Cummings model (JCM) \cite{jaynes63}, which is valid when the detuning between the qubit transition frequency $\Omega$ and the resonator frequency $\omega$ is negligible $(\Omega \approx \omega)$ and the coupling between the qubit and oscillator is weak $(g<\omega)$ \cite{kok10}. This is an excellent approximation in the case of cavity QED where typical coupling strengths are of order $g/\omega\approx10^{-6}$. 
The JCM can be extended to incorporate tunnelling, 
and has provided an adequate description of experiments for decades, but a new era of experiments are pushing beyond its boundaries in terms of both coupling and detuning. Circuit QED experiments  couple superconducting qubits to LC and waveguide resonators, allowing coupling strengths up to $g/\omega\approx10^{-1}$, recently enabling demonstrations of the breakdown of the JCM  \cite{niemczyk2010, forn-diaz2010}. Superconducting qubits coupled to nanomechanical resonators (NR) generally have more modest coupling strengths  \cite{LaHaye09, OConnell10}, but combined with large detuning they could also operate outside the validity of the JCM \cite{irish05}.

The Hamiltonian for the Rabi model can be decomposed into three parts:
\begin{equation}\label{eqn:hamBreak}
\hat{H}= \hat{H}_{Q}+ \hat{H}_{O}+\hat{H}_{I}.
\end{equation}
The qubit, atom or two level system is described by: 
\begin{equation}\label{eqn:hamQ}
\hat{H}_{Q}= \frac{\epsilon}{2}\sigma_{z}+\frac{\Delta}{2}\sigma_{x},
\end{equation}
where $\sigma_{z}$ and $\sigma_{x}$ are the Pauli spin operators. They describe a two level system with an energy splitting $\epsilon$ and a spontaneous tunnelling between the states at a rate $\Delta$. In isolation such a system would undergo Rabi oscillations with a frequency $\Omega_{r}=\sqrt{\epsilon^{2}+\Delta^{2}}$. 
The Hamiltonian of the oscillator is: 
\begin{equation}\label{eqn:hamO}
\hat{H}_{O}=  \omega a^{\dagger}a,
\end{equation}
where  $\omega$ is the frequency of the oscillator and  $a^{\dagger}$ and $a$ are its creation and annihilation operators respectively. Note we have neglected the zero point energy. The Hamiltonian for the interaction between the two is:
\begin{equation}\label{eqn:hamI}
\hat{H}_{I}=  g(a+a^{\dagger})\sigma_{z},
\end{equation}
where $g$ is the coupling strength between the qubit and oscillator. 

Recent experimental progress has sparked a renewed theoretical interest in extending solutions of (\ref{eqn:hamBreak}) beyond the RWA. For instance, a change of basis prior to applying the RWA leads to a  generalised RWA that should be valid outside the very weak coupling limit \cite{irish07}. However, this is limited to the case of $\epsilon=0$. As an alternative approach, Van Vleck perturbation theory \cite{vanvleck29} has been used to investigate the dynamics in the ultra strong ($g/\omega > 1$) coupling regime \cite{hausinger10,hausinger11}. This approach contains the splitting and tunnelling elements, but it is perturbative in the latter and fails to recover the JCM in the weak coupling limit. This approach is therefore more applicable to circuit QED, rather than the more modest couplings achieved in Cooper pair box (CPB) coupled to NR systems. 

An analytic expression for the eigenspectrum of the full Rabi model was very recently found by Braak \cite{braak11},  a surprising and significant result for such a long standing problem. However, it is too early to tell how much this solution can reveal about the physical properties and dynamics of the system. In addition to solving the model Braak proved that it is non-integrable i.e. the time-dependence of important properties cannot be found in closed form. There is therefore still a need for approximate results governing areas of particular experimental interest.

In order to simplify the expression and extend the validity of the approximations that we will subsequently describe into the strong coupling regime, we first perform a `polaron' transformation \cite{wagner86, mahan00}. This unitary Hamiltonian transformation ($H'=e^{s}He^{-s}$) is equivalent to dressing qubit excitations with the vibrational modes to form quasi-particles called polarons.  With $s = \alpha /2 (\cre - \an) \sz$ and $\alpha/2 = g / \omega$  we obtain
\begin{equation}\label{eqn:hamPolar}
H'  = \frac{\epsilon}{2} \sigma_{z} + \omega \no + \frac{\Delta}{2} (D(\alpha) \kb{0}{1} + D(-\alpha) \kb{1}{0}),
\end{equation}
where $D(\xi) = \exp (\xi \cre - \xi^* \an )$ is the displacement operator. We have neglected a term proportional to the identity $g^2 / \omega~\id$, which does not influence the dynamics. The first two terms involve the qubit and oscillator individually and so can be removed by going to the interaction picture. We insert the resulting Hamiltonian into the von Neumann equation and then derive equations of motion for the qubit (see Appendix and Refs.~\cite{brandes03,brandes05}):
\begin{eqnarray}
&\ddt \rho_{00}(t)   =& -i\frac{\Delta}{2} (\rho_{10}(t) - \rho_{01}(t))), \label{eqn:psol1}\\
&\ddt \rho_{11}(t)   =&\phantom{-} i\frac{\Delta}{2} (\rho_{10}(t) - \rho_{01}(t))), \label{eqn:psol2}
\end{eqnarray}
where $ \rho_{00}(t)$ and  $\rho_{11}(t)$ are the time dependant population elements of the qubit's reduced density matrix. The coherences are given by: 
\begin{eqnarray}
\rho_{01}(t) & = i\frac{\Delta}{2}  \int_0^t \mathrm{d}t'  e^{-i \epsilon\tau} \left[ \rho_{00}(t') C^*(\tau)-\rho_{11}(t') C(\tau)\right], \label{eqn:psol3} \\
\rho_{10}(t)   &= -i\frac{\Delta}{2}  \int_0^t \mathrm{d}t' e^{i \epsilon\tau} \left[ \rho_{00}(t') C(\tau)-\rho_{11}(t') C^*(\tau)   \right],  \label{eqn:psol4}
\end{eqnarray}
where $\tau=t-t'$ and $C(\tau)$ and $C^*(\tau)$ are the correlation function of the oscillator and its complex conjugate respectively. In deriving these equations, we have employed the Born approximation, i.e. we have assumed that the vibrational mode and the qubit states can be factored at all times.

Physically, this corresponds to an oscillator that thermalises on a timescale faster than that characteristic of the qubit dynamics. 

The equations of motion take the form of a system of integro-differential equations involving the  bosonic correlation function and its complex conjugate. Laplace transforming the equations of motion yields a set of simultaneous equations that can be solved algebraically: 
\begin{eqnarray}
&R_{00}(s)& = \frac{s\rho_{0} + \left(\frac{\Delta}{2} \right)^{2} \left[ C'_{+} + C''_{-} \right] }{s^{2}+ s \left( \frac{\Delta}{2} \right)^{2} \left[C'_{-}+C''_{-}+C'_{+}+C''_{+} \right]}\label{eqn:rsol1}\\
&R_{10}(s)& =  -i \frac{\Delta}{2} \left[ (C'_{-} + C''_{-})R_{00}(s) -\frac{1}{s}C''_{-} \right] \label{eqn:rsol2}
\end{eqnarray}
where $s$ is our Laplace space variable,  $R_{00}(s)$ and $R_{10}(s)$ are the Laplace transforms of $\rho_{00}(t)$ and $\rho_{10}(t)$, $\rho_{0}$ is the initial population of the ground state and $C'_{\pm}=C'(s\pm i\epsilon)$ and $C''_{\pm}$ are the Laplace transforms of the correlation function and its conjugate respectively. It is sufficient to solve these two equations alone because from their solutions the behaviour of the other density matrix elements can be trivially derived. 

To obtain expressions for the dynamics of Eqns (\ref{eqn:rsol1}) and (\ref{eqn:rsol2}) in the time domain we need to find the Laplace transform of the bosonic correlation and its conjugate, solve and then take the inverse Laplace transform of the equations. The correlation function is defined as
\begin{equation} \label{eqn:bosoncf}
C(\tau) = \exx{D_{t}(\alpha) D_{t'}^{\dagger}(\alpha)} =   \text{Tr}_B \left[ \rho_B D_t(\alpha) D_{t'}^{\dagger}(\alpha) \right] 
\end{equation}
which evaluates to (see Appendix):
\begin{equation}\label{eqn:mah}
C(\tau) = e^{-|\alpha|^{2}((1-\cos{(\omega \tau)})\coth{\frac{\beta\omega}{2}} +i\sin{(\omega \tau)})}.
\end{equation}
Unfortunately, it is not straightforward to Laplace transform this expression directly, so we employ the Jacobi-Anger series expansion:
\begin{equation}\label{JA}
e^{z\cos{\theta}} = \sum_{n=-\infty}^{\infty} I_n(z) e^{in\theta},
\end{equation} 
where $z$ is an arbitrary complex number and $I_n(z)$ is the modified Bessel function of order $n$ and argument $z$. By exploiting an angle addition identity we can rewrite (\ref{eqn:mah}) as:
\begin{equation}\label{Sin4}
C(\tau) = e^{- |\alpha|^{2} \coth{(\frac{\beta\omega}{2})}}e^{z\cos{(\omega \tau +x)}},
\end{equation}

where $x=i\beta\omega / 2$ and $z= 2 |\alpha|^{2}\sqrt{N(N+1)}$. Using Eqn~(\ref{JA}) this gives:
\begin{equation}\label{Sin5}
C(\tau) =  e^{-|\alpha|^{2}(2N+1)}\sum _{n=-\infty}^{\infty}I_n(z)e^{i n ( \omega  \tau +x)};
\end{equation}
where $N=(e^{\beta \omega}-1)^{-1}$ is the average oscillator occupation number. In this form the correlation function can be Laplace transformed trivially. The Bessel function weighting of the series means it converges very rapidly, in fact only a few terms of the series need to be retained to accurately capture the dynamics. For experimentally relevant parameters (i.e.~low temperatures and moderate to strong coupling) only a single term dominates. The corresponds to the regime where:
\begin{equation}\label{eqn:crit}
 2|\alpha|^{2} \sqrt{N(N+1)} \ll 1.
\end{equation}

%Figure
\begin{figure}
\includegraphics[scale=0.30]{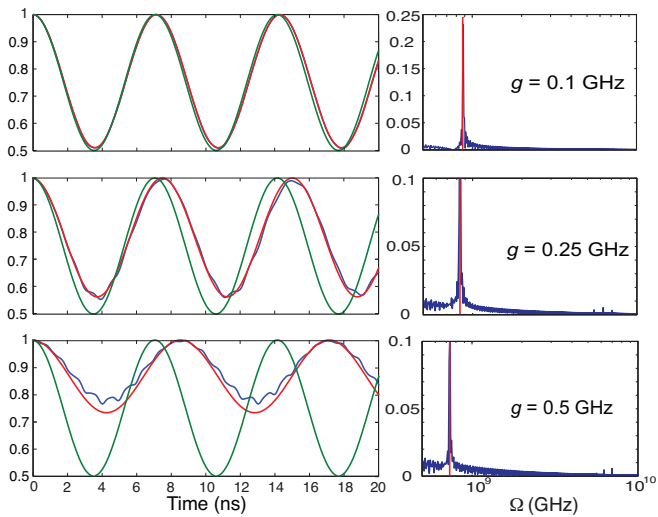}
\caption{\label{fig:1} Comparison of the single term approximation (red) and a numerically exact approach (blue) for different coupling strengths. Uncoupled Rabi oscillations are also shown as a reference (green). Left: the population $\rho_{00}(t)$ in the time-domain.  Right:  the same data in the frequency domain. The full numerical solution was Fourier transformed using Matlab's FFT algorithm.  Other parameters are $\omega = 1$ GHz, $g=0.1$ GHz, $\epsilon = \Delta = 100$ MHz and $T = 10$ mK.}
\end{figure}
%Figure end

Physically this corresponds to retaining only interactions that conserve the total phonon number in the oscillator, thus complementing the underlying Born Approximation, which assumes the oscillator remains in thermal equilibrium. Including only the dominant zeroth term in the series allows the equations (\ref{eqn:rsol1}) and (\ref{eqn:rsol2}) to be inverse Laplace transformed: 
\begin{equation}\label{eqn:rho0}
\rho_{00}(t) =\frac{\rho _0 \epsilon ^2+\frac{1}{2} e^{-b} \Delta ^2 I_0(z) \left(\left(2 \rho _0-1\right) \cos (t \Omega )+1\right)}{\Omega ^2},
\end{equation}
\begin{equation}\label{eqn:rho1}
\rho_{10}(t) =-\frac{e^{-b} \Delta  \left(2 \rho _0-1\right) I_0(z) (\epsilon  \cos (t \Omega )+i \Omega  \sin (t \Omega )-\epsilon )}{2 \Omega ^2},
\end{equation}
\begin{equation}\label{eqn:rho3}
\Omega =  \sqrt{ \Delta ^2e^{-b}I_0(z)+\epsilon ^2},
\end{equation}
\newline
where $b=|\alpha|^{2} (2N +1)$. From Eqn (\ref{eqn:rho3}) we can see that the presence of the oscillator alters the tunnelling rate by a factor $I_0(z)e^{-b}$, essentially corresponding to a temperature dependent ac Stark shift on the qubit due to the presence of the single oscillator mode \cite{irish03, schuster05}. In contrast to previous work our expression is not confined to the weak coupling or large detunning limit, but still takes a surprisingly simple closed form.

Figure \ref{fig:1} shows a comparison of the dynamics predicted using these expressions and a numerically exact approach. The latter are obtained by imposing a truncation of the oscillator Hilbert space at a point where the dynamics have converged and any higher modes have an extremely low occupation probability. Our zeroth order approximation proves to be unexpectedly powerful, giving accurate dynamics well into the strong coupling regime ($g/\omega=0.25$) and even beyond this it still captures the dominant oscillatory behaviour, see Figure \ref{fig:1}. Stronger coupling increases the numerical weight of higher frequency terms in the series, causing a modulation of the dynamics. The approximation starts to break down at ($g/\omega=0.5$). The equations (\ref{eqn:rho0}) and (\ref{eqn:rho1}) are obviously unable to capture the higher frequency modulations to the dynamics or any potential long time phenomena like collapse and revival, but these are unlikely to be resolvable in experiments in any case.  Nonetheless, it is worth pointing out that even in this strong coupling case the base frequency of the qubit dynamics is still adequately captured by our single term approximation.

Our methodology can be used to predict dynamics of nanomechanical resonators connected to either quantum dots or superconducting qubits. The criterion for the single term approximation to be valid is readily met by current experiments such as those presented in Refs.~\cite{LaHaye09, OConnell10} and their parameters yield near perfect agreement between numerical and analytic results. Most experiments operate in a regime where the qubit dynamics are not greatly perturbed by the presence of the oscillator, which has a much lower frequency ($\epsilon \approx \Delta \approx 10$ GHz, $\omega =1$ GHz).  In Figure \ref{fig:1}, we chose $\epsilon \approx \Delta \approx 100$ MHz, because this better demonstrates the effect of the oscillator on the qubit. These parameters can be achieved experimentally using the same qubit design but with an oscillating voltage applied to the CPB bias gate \cite{irish05}. However, we stress the accuracy of our method is not restricted to this regime. 

%Figure
\begin{figure}
\includegraphics[scale=1]{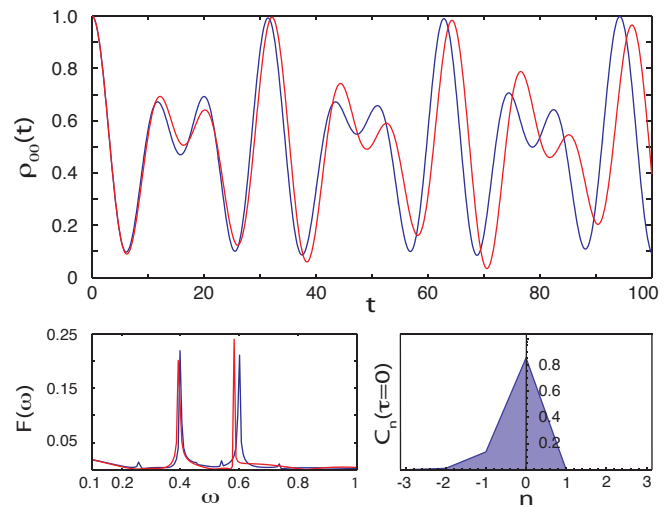}
\caption{\label{fig:2} Main panel: comparison of dynamics calculated from truncating (\ref{Sin5}) at $N_{\rm MAX}=\pm10$ (red) and a numerically exact approach (blue). Lower left:  Fourier transform of the dynamics.  Lower right:  the numerical weight of the $n^{th}$ term in the series expansion of (\ref{Sin5}), showing there are still only two dominant frequencies at $n=0$ and $n=-1$. Parameters:  $\omega = 0.5$, $g=0.1$, $\epsilon =0 $, $ \Delta = 0.5$, $T = 1~\rm{mK}$, $\hbar=1$ and $k_{b}=1$.}
\end{figure}
%Figure end

Including extra terms in the series expansion (\ref{Sin5}) makes the time dependence of the qubit dynamics analytically unwieldy, because the rational function form of the series leads to a complex interdepence of the positions of the poles in (\ref{eqn:rsol1}). However, if the values of the parameters are known the series can truncated at ($\pm N_{\rm MAX}$) to give an efficient numerical method to obtain more accurate dynamics, extending the applicability of our approach beyond the regime described by (\ref{eqn:crit}). This is demonstrated in Fig. \ref{fig:2}, where the dynamics are clearly dominated by two frequencies -- an effect that could obviously never be captured by a single term approximation. There is a qualitative agreement between the many terms expansion and full numerical solution, particularly at short times. We would not expect a perfect agreement in this case because the simulations are of the dynamics in the large tunnelling regime ($\Delta=0.5$), and the polaron transform makes the master equation perturbative in this parameter.  Nonetheless, the rapid convergence of the series is shown in Fig. \ref{fig:2}; $N_{\rm MAX}=5-10$ is sufficient to calculate $\rho_{00}(t)$ and $\rho_{10}(t)$ with an accuracy only limited by the underlying Born Approximation. The asymmetry of the amplitudes of the terms in the series expansion of (\ref{Sin5}) is due to the exponential functions in the series.

%\section{Quantum Thermometry}

%Figure
\begin{figure}
\includegraphics[scale=1]{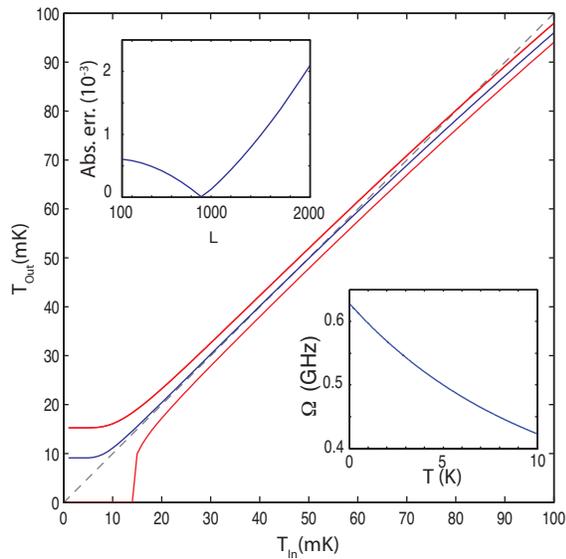}
\caption{\label{fig:3} Demonstration of qubit thermometry: $T_{\rm in}$ is the temperature supplied to the numerical simulation of the system and $T_{\rm out}$ is the temperature that would be predicted by fitting oscillations with frequency (\ref{eqn:rho3}) to it. The blue line is the data and red line shows the effect of a 10kHz error in the frequency measurement; the grey dashed line serves as a guide to the eye. The lower inset shows the variation of the qubit frequency $\Omega$ with temperature. The upper inset shows the dependence of the absolute error in the prediction against the signal length (see text). Other parameters are: $\omega = 1$ GHz, $g=0.01$ GHz, $\epsilon =0$,  $\Delta = 100$ MHz} 
\end{figure}
%Figure end

We now discuss the application of our model system to the measurement of the temperature of an oscillator by observing the coupled qubit. We picture a situation in which the qubit oscillation frequency $\Omega$ is the measured quantity. The tunnelling, coupling strength and energy splitting are usually within the control of the experimentalist (or are at least known constants), and this yields the possibility of using the measured $\Omega$ to estimate the temperature. A related idea was recently used in the calibration of a seminal resonator experiment \cite{OConnell10} to verify that the oscillator was in its ground state (a critical part of the work). In that case, the authors used a comparison of numerical results for different occupation numbers $N$ with the measured population in the excited state of the qubit after a certain interaction time.  A theoretical study of the same approach was preformed in \cite{Brunelli11}, where the system was described by the JCM without a tunnelling term. In contrast, we here propose that simple analytic expressions for the qubit dynamics that are valid beyond the weak coupling regime, like our own, can be used to directly measure the temperature and hence $N$ of the oscillator, simply by observing the effective qubit Rabi frequency $\Omega$. 

Figure \ref{fig:3} demonstrates this idea, showing that by measuring $\Omega$ and fitting it to our expression (\ref{eqn:rho3}), we can obtain submilli-Kelvin precision in the experimentally relevant regime of 20-55 mK. At low temperatures the single term frequency plateaus, causing the accuracy to break down. In the higher temperature limit, we also see a deviation from the diagonal, this is to be expected as we leave the regime of validity described by (\ref{eqn:crit}). Naturally accuracy in this region could be improved by retaining higher order terms in (\ref{Sin5}), but this would become a more numeric than analytic approach. The upper inset shows the dependence of the accuracy of the prediction on the number of points (at a separation of 1ns) sampled from the dynamics. The accuracy increases initially as more points improve the fitted value of $\Omega$, however after a certain length the accuracy is diminished by long term envelope effects in the dynamics not captured by the single term approximation. We note that the corresponding analysis in the frequency domain would not be equally affected by the long time envelope, however a large number of points in the FFT is then required in order to obtain the desired accuracy. The lower inset of Figure \ref{fig:3} shows the direct dependence of $\Omega$ on the temperature. The temperature range with steepest gradient and hence greatest frequency dependence on temperature varies with the coupling strength; thus the device could be specifically designed to have a maximal sensitivity in the temperature range of the most interest. 

In summary, we have developed and explored a new approach to the qubit-oscillator model, finding succinct expressions for the qubit dynamics. In contrast to previous theoretical approaches, our expressions are valid in the stronger coupling regime that is rapidly gaining experimental relevance. We have further proposed an application of our model enabling precise temperature measurements of the oscillator mode. This could be used either as part of the calibration of an oscillator experiment or as a tuned, standalone device. 

\begin{acknowledgments}
We thank Nikesh Dattani, Tom Stace, Gerard Milburn, and Ahsan Nazir for fruitful discussions. This work was supported by the EPSRC, the National Research Foundation and Ministry of Education, Singapore, and the Royal Society.
\end{acknowledgments}

\bibliographystyle{apsrev}

\begin{thebibliography}{25}
\expandafter\ifx\csname natexlab\endcsname\relax\def\natexlab#1{#1}\fi
\expandafter\ifx\csname bibnamefont\endcsname\relax
  \def\bibnamefont#1{#1}\fi
\expandafter\ifx\csname bibfnamefont\endcsname\relax
  \def\bibfnamefont#1{#1}\fi
\expandafter\ifx\csname citenamefont\endcsname\relax
  \def\citenamefont#1{#1}\fi
\expandafter\ifx\csname url\endcsname\relax
  \def\url#1{\texttt{#1}}\fi
\expandafter\ifx\csname urlprefix\endcsname\relax\def\urlprefix{URL }\fi
\providecommand{\bibinfo}[2]{#2}
\providecommand{\eprint}[2][]{\url{#2}}

\bibitem[{\citenamefont{Rabi}(1936)}]{rabi36}
\bibinfo{author}{\bibfnamefont{I.~I.} \bibnamefont{Rabi}},
  \bibinfo{journal}{Phys. Rev.} \textbf{\bibinfo{volume}{49}},
  \bibinfo{pages}{324} (\bibinfo{year}{1936}).

\bibitem[{\citenamefont{Shore and Knight}(1993)}]{shore1993}
\bibinfo{author}{\bibfnamefont{B.~W.} \bibnamefont{Shore}} \bibnamefont{and}
  \bibinfo{author}{\bibfnamefont{P.~L.} \bibnamefont{Knight}},
  \bibinfo{journal}{Journal of Modern Optics} \textbf{\bibinfo{volume}{40}},
  \bibinfo{pages}{1195} (\bibinfo{year}{1993}).

\bibitem[{\citenamefont{Raimond et~al.}(2001)\citenamefont{Raimond, Brune, and
  Haroche}}]{raimond01}
\bibinfo{author}{\bibfnamefont{J.~M.} \bibnamefont{Raimond}},
  \bibinfo{author}{\bibfnamefont{M.}~\bibnamefont{Brune}}, \bibnamefont{and}
  \bibinfo{author}{\bibfnamefont{S.}~\bibnamefont{Haroche}},
  \bibinfo{journal}{Rev. Mod. Phys.} \textbf{\bibinfo{volume}{73}},
  \bibinfo{pages}{565} (\bibinfo{year}{2001}).

\bibitem[{\citenamefont{Fulton and Gouterman}(1961)}]{fulton61}
\bibinfo{author}{\bibfnamefont{R.~L.} \bibnamefont{Fulton}} \bibnamefont{and}
  \bibinfo{author}{\bibfnamefont{M.}~\bibnamefont{Gouterman}},
  \bibinfo{journal}{Journal of Chemical Physics} \textbf{\bibinfo{volume}{35}},
  \bibinfo{pages}{1059} (\bibinfo{year}{1961}), ISSN \bibinfo{issn}{00219606}.

\bibitem[{\citenamefont{Jaynes and Cummings}(1963)}]{jaynes63}
\bibinfo{author}{\bibfnamefont{E.}~\bibnamefont{Jaynes}} \bibnamefont{and}
  \bibinfo{author}{\bibfnamefont{F.}~\bibnamefont{Cummings}},
  \bibinfo{journal}{Proceedings of the IEEE} \textbf{\bibinfo{volume}{51}},
  \bibinfo{pages}{89 } (\bibinfo{year}{1963}), ISSN \bibinfo{issn}{0018-9219}.

\bibitem[{\citenamefont{Kok and Lovett}(2010)}]{kok10}
\bibinfo{author}{\bibfnamefont{P.}~\bibnamefont{Kok}} \bibnamefont{and}
  \bibinfo{author}{\bibfnamefont{B.}~\bibnamefont{Lovett}},
  \emph{\bibinfo{title}{Optical Quantum Information Processing}}
  (\bibinfo{publisher}{Cambridge University Press}, \bibinfo{year}{2010}).

\bibitem[{\citenamefont{Niemczyk et~al.}(2010)\citenamefont{Niemczyk, Deppe,
  Huebl, Menzel, Hocke, Schwarz, Garcia-Ripoll, Zueco, Hummer, Solano
  et~al.}}]{niemczyk2010}
\bibinfo{author}{\bibfnamefont{T.}~\bibnamefont{Niemczyk}},
  \bibinfo{author}{\bibfnamefont{F.}~\bibnamefont{Deppe}},
  \bibinfo{author}{\bibfnamefont{H.}~\bibnamefont{Huebl}},
  \bibinfo{author}{\bibfnamefont{E.~P.} \bibnamefont{Menzel}},
  \bibinfo{author}{\bibfnamefont{F.}~\bibnamefont{Hocke}},
  \bibinfo{author}{\bibfnamefont{M.~J.} \bibnamefont{Schwarz}},
  \bibinfo{author}{\bibfnamefont{J.~J.} \bibnamefont{Garcia-Ripoll}},
  \bibinfo{author}{\bibfnamefont{D.}~\bibnamefont{Zueco}},
  \bibinfo{author}{\bibfnamefont{T.}~\bibnamefont{Hummer}},
  \bibinfo{author}{\bibfnamefont{E.}~\bibnamefont{Solano}},
  \bibnamefont{et~al.}, \bibinfo{journal}{Nat Phys}
  \textbf{\bibinfo{volume}{6}}, \bibinfo{pages}{772} (\bibinfo{year}{2010}).

\bibitem[{\citenamefont{Forn-D\'iaz et~al.}(2010)\citenamefont{Forn-D\'iaz,
  Lisenfeld, Marcos, Garc\'ia-Ripoll, Solano, Harmans, and
  Mooij}}]{forn-diaz2010}
\bibinfo{author}{\bibfnamefont{P.}~\bibnamefont{Forn-D\'iaz}},
  \bibinfo{author}{\bibfnamefont{J.}~\bibnamefont{Lisenfeld}},
  \bibinfo{author}{\bibfnamefont{D.}~\bibnamefont{Marcos}},
  \bibinfo{author}{\bibfnamefont{J.~J.} \bibnamefont{Garc\'ia-Ripoll}},
  \bibinfo{author}{\bibfnamefont{E.}~\bibnamefont{Solano}},
  \bibinfo{author}{\bibfnamefont{C.~J. P.~M.} \bibnamefont{Harmans}},
  \bibnamefont{and} \bibinfo{author}{\bibfnamefont{J.~E.} \bibnamefont{Mooij}},
  \bibinfo{journal}{Phys. Rev. Lett.} \textbf{\bibinfo{volume}{105}},
  \bibinfo{pages}{237001} (\bibinfo{year}{2010}).

\bibitem[{\citenamefont{LaHaye et~al.}(2009)\citenamefont{LaHaye, Suh,
  Echternach, Schwab, and Roukes}}]{LaHaye09}
\bibinfo{author}{\bibfnamefont{M.~D.} \bibnamefont{LaHaye}},
  \bibinfo{author}{\bibfnamefont{J.}~\bibnamefont{Suh}},
  \bibinfo{author}{\bibfnamefont{P.~M.} \bibnamefont{Echternach}},
  \bibinfo{author}{\bibfnamefont{K.~C.} \bibnamefont{Schwab}},
  \bibnamefont{and} \bibinfo{author}{\bibfnamefont{M.~L.}
  \bibnamefont{Roukes}}, \bibinfo{journal}{Nature}
  \textbf{\bibinfo{volume}{459}}, \bibinfo{pages}{960} (\bibinfo{year}{2009}).

\bibitem[{\citenamefont{O'Connell et~al.}(2010)\citenamefont{O'Connell,
  Hofheinz, Ansmann, Bialczak, Lenander, Lucero, Neeley, Sank, Wang, Weides
  et~al.}}]{OConnell10}
\bibinfo{author}{\bibfnamefont{A.~D.} \bibnamefont{O'Connell}},
  \bibinfo{author}{\bibfnamefont{M.}~\bibnamefont{Hofheinz}},
  \bibinfo{author}{\bibfnamefont{M.}~\bibnamefont{Ansmann}},
  \bibinfo{author}{\bibfnamefont{R.~C.} \bibnamefont{Bialczak}},
  \bibinfo{author}{\bibfnamefont{M.}~\bibnamefont{Lenander}},
  \bibinfo{author}{\bibfnamefont{E.}~\bibnamefont{Lucero}},
  \bibinfo{author}{\bibfnamefont{M.}~\bibnamefont{Neeley}},
  \bibinfo{author}{\bibfnamefont{D.}~\bibnamefont{Sank}},
  \bibinfo{author}{\bibfnamefont{H.}~\bibnamefont{Wang}},
  \bibinfo{author}{\bibfnamefont{M.}~\bibnamefont{Weides}},
  \bibnamefont{et~al.}, \bibinfo{journal}{Nature}
  \textbf{\bibinfo{volume}{464}}, \bibinfo{pages}{697} (\bibinfo{year}{2010}).

\bibitem[{\citenamefont{Irish et~al.}(2005)\citenamefont{Irish, Gea-Banacloche,
  Martin, and Schwab}}]{irish05}
\bibinfo{author}{\bibfnamefont{E.~K.} \bibnamefont{Irish}},
  \bibinfo{author}{\bibfnamefont{J.}~\bibnamefont{Gea-Banacloche}},
  \bibinfo{author}{\bibfnamefont{I.}~\bibnamefont{Martin}}, \bibnamefont{and}
  \bibinfo{author}{\bibfnamefont{K.~C.} \bibnamefont{Schwab}},
  \bibinfo{journal}{Phys. Rev. B} \textbf{\bibinfo{volume}{72}},
  \bibinfo{pages}{195410} (\bibinfo{year}{2005}).

\bibitem[{\citenamefont{Irish}(2007)}]{irish07}
\bibinfo{author}{\bibfnamefont{E.~K.} \bibnamefont{Irish}},
  \bibinfo{journal}{Phys. Rev. Lett.} \textbf{\bibinfo{volume}{99}},
  \bibinfo{pages}{173601} (\bibinfo{year}{2007}).

\bibitem[{\citenamefont{Van~Vleck}(1929)}]{vanvleck29}
\bibinfo{author}{\bibfnamefont{J.~H.} \bibnamefont{Van~Vleck}},
  \bibinfo{journal}{Phys. Rev.} \textbf{\bibinfo{volume}{33}},
  \bibinfo{pages}{467} (\bibinfo{year}{1929}).

\bibitem[{\citenamefont{Hausinger and Grifoni}(2010)}]{hausinger10}
\bibinfo{author}{\bibfnamefont{J.}~\bibnamefont{Hausinger}} \bibnamefont{and}
  \bibinfo{author}{\bibfnamefont{M.}~\bibnamefont{Grifoni}},
  \bibinfo{journal}{Phys. Rev. A} \textbf{\bibinfo{volume}{82}},
  \bibinfo{pages}{062320} (\bibinfo{year}{2010}).

\bibitem[{\citenamefont{Hausinger and Grifoni}(2011)}]{hausinger11}
\bibinfo{author}{\bibfnamefont{J.}~\bibnamefont{Hausinger}} \bibnamefont{and}
  \bibinfo{author}{\bibfnamefont{M.}~\bibnamefont{Grifoni}},
  \bibinfo{journal}{Phys. Rev. A} \textbf{\bibinfo{volume}{83}},
  \bibinfo{pages}{030301} (\bibinfo{year}{2011}).

\bibitem[{\citenamefont{Braak}(2011)}]{braak11}
\bibinfo{author}{\bibfnamefont{D.}~\bibnamefont{Braak}},
  \bibinfo{journal}{Phys. Rev. Lett.} \textbf{\bibinfo{volume}{107}},
  \bibinfo{pages}{100401} (\bibinfo{year}{2011}).

\bibitem[{\citenamefont{Wagner}(1986)}]{wagner86}
\bibinfo{author}{\bibfnamefont{M.}~\bibnamefont{Wagner}},
  \emph{\bibinfo{title}{Unitary Transforms in Solid State Physics}}
  (\bibinfo{publisher}{North-Holland}, \bibinfo{year}{1986}),
  \bibinfo{edition}{1st} ed.

\bibitem[{\citenamefont{Mahan}(2000)}]{mahan00}
\bibinfo{author}{\bibfnamefont{G.~D.} \bibnamefont{Mahan}},
  \emph{\bibinfo{title}{Many Particle Physics (Physics of Solids and Liquids)}}
  (\bibinfo{publisher}{Springer}, \bibinfo{year}{2000}), \bibinfo{edition}{3rd}
  ed., ISBN \bibinfo{isbn}{0306463385}.

\bibitem[{\citenamefont{Brandes and Lambert}(2003)}]{brandes03}
\bibinfo{author}{\bibfnamefont{T.}~\bibnamefont{Brandes}} \bibnamefont{and}
  \bibinfo{author}{\bibfnamefont{N.}~\bibnamefont{Lambert}},
  \bibinfo{journal}{Phys. Rev. B} \textbf{\bibinfo{volume}{67}},
  \bibinfo{pages}{125323} (\bibinfo{year}{2003}).

\bibitem[{\citenamefont{Brandes}(2005)}]{brandes05}
\bibinfo{author}{\bibfnamefont{T.}~\bibnamefont{Brandes}},
  \bibinfo{journal}{Physics Reports} \textbf{\bibinfo{volume}{408}},
  \bibinfo{pages}{315} (\bibinfo{year}{2005}).

\bibitem[{\citenamefont{Irish and Schwab}(2003)}]{irish03}
\bibinfo{author}{\bibfnamefont{E.~K.} \bibnamefont{Irish}} \bibnamefont{and}
  \bibinfo{author}{\bibfnamefont{K.}~\bibnamefont{Schwab}},
  \bibinfo{journal}{Phys. Rev. B} \textbf{\bibinfo{volume}{68}},
  \bibinfo{pages}{155311} (\bibinfo{year}{2003}).

\bibitem[{\citenamefont{Schuster et~al.}(2005)\citenamefont{Schuster, Wallraff,
  Blais, Frunzio, Huang, Majer, Girvin, and Schoelkopf}}]{schuster05}
\bibinfo{author}{\bibfnamefont{D.~I.} \bibnamefont{Schuster}},
  \bibinfo{author}{\bibfnamefont{A.}~\bibnamefont{Wallraff}},
  \bibinfo{author}{\bibfnamefont{A.}~\bibnamefont{Blais}},
  \bibinfo{author}{\bibfnamefont{L.}~\bibnamefont{Frunzio}},
  \bibinfo{author}{\bibfnamefont{R.-S.} \bibnamefont{Huang}},
  \bibinfo{author}{\bibfnamefont{J.}~\bibnamefont{Majer}},
  \bibinfo{author}{\bibfnamefont{S.~M.} \bibnamefont{Girvin}},
  \bibnamefont{and} \bibinfo{author}{\bibfnamefont{R.~J.}
  \bibnamefont{Schoelkopf}}, \bibinfo{journal}{Phys. Rev. Lett.}
  \textbf{\bibinfo{volume}{94}}, \bibinfo{pages}{123602}
  (\bibinfo{year}{2005}).

\bibitem[{\citenamefont{Brunelli et~al.}(2011)\citenamefont{Brunelli, Olivares,
  and Paris}}]{Brunelli11}
\bibinfo{author}{\bibfnamefont{M.}~\bibnamefont{Brunelli}},
  \bibinfo{author}{\bibfnamefont{S.}~\bibnamefont{Olivares}}, \bibnamefont{and}
  \bibinfo{author}{\bibfnamefont{M.~G.~A.} \bibnamefont{Paris}},
  \bibinfo{journal}{arXiv}  (\bibinfo{year}{2011}), \eprint{1103.2875}.

\bibitem[{\citenamefont{de~Oliveira et~al.}(2005)\citenamefont{de~Oliveira,
  de~Almeida, de~Queiros, Moraes, and Dantas}}]{oliveira05}
\bibinfo{author}{\bibfnamefont{G.~C.} \bibnamefont{de~Oliveira}},
  \bibinfo{author}{\bibfnamefont{A.~R.} \bibnamefont{de~Almeida}},
  \bibinfo{author}{\bibfnamefont{I.~P.} \bibnamefont{de~Queiros}},
  \bibinfo{author}{\bibfnamefont{A.~M.} \bibnamefont{Moraes}},
  \bibnamefont{and} \bibinfo{author}{\bibfnamefont{C.~M.~A.}
  \bibnamefont{Dantas}}, \bibinfo{journal}{Physica A: Statistical Mechanics and
  its Applications} \textbf{\bibinfo{volume}{351}}, \bibinfo{pages}{251}
  (\bibinfo{year}{2005}).

\bibitem[{\citenamefont{Crisp}(1992)}]{crisp92}
\bibinfo{author}{\bibfnamefont{M.~D.} \bibnamefont{Crisp}},
  \bibinfo{journal}{Phys. Rev. A} \textbf{\bibinfo{volume}{46}},
  \bibinfo{pages}{4138} (\bibinfo{year}{1992}).

\end{thebibliography}

\onecolumngrid
\appendix

\section{Appendix: derivation of the equations of motion and the bosonic correlation function}

In this Appendix we give the explicit derivation of the equations of motion (\ref{eqn:psol1}-\ref{eqn:psol4}), the Laplace transformed equations of motion  (\ref{eqn:rsol1},\ref{eqn:rsol2}), and the bosonic correlation function (\ref{eqn:bosoncf}). We note that parts of these derivations can be found in similar form in the literature (cf. Refs.~\cite{brandes03, brandes05, mahan00}), but we here give an alternate and full account in consistent notation for the benefit of the reader.

\subsection{ \label{app:a} Equations of Motion}
First we move into the interaction picture:

\begin{eqnarray}
\tilde{\rho}_{00}(t)= \rho_{00},~~
\tilde{\rho}_{11}(t)= \rho_{11},~~
\tilde{\rho}_{01}(t)= \rho_{01}e^{i\epsilon t}D_t ,~~
\tilde{\rho}_{10}(t)= \rho_{10}e^{-i\epsilon t}D^{\dagger}_t,
\end{eqnarray}
where $D_t$ and $D^{\dagger}_t$ are the time dependent versions of the displacement operators introduced by the polaron transform. 
\begin{equation}
\tilde{H}_I\phantom{'}(t)=  \frac{\Delta}{2} \left( \rho_{01}(t)  +  \rho_{10}(t)  \right), 
\end{equation}

Starting from the Von Neumann equation:
\begin{eqnarray}
&\frac{d}{dt}\tilde{\rho }(t)=-i\left[\tilde{H}_I(t),\tilde{\rho }(t)\right], \\
&\tilde{\rho }(t)=\rho_{0}-i\int_0^t{dt'[\tilde{H}_I(t'),\tilde{\rho }(t')]}, \label{neuint}
\end{eqnarray}

To study the dynamics we need the time dependent expectation values of the density matrix elements. These are given by:

\begin{equation}\label{expect1}
\exx{O}_t=\text{Tr}[\rho(t)O]= \text{Tr}[\tilde{\rho}(t)\tilde{O}_t],
\end{equation}

Substituting in (\ref{neuint}) 
\begin{equation}\label{expect2}
\exx{O}_t -\exx{O}_0 =-i\int_0^t{dt' \text{Tr}[[\tilde{H}_I(t'),\tilde{\rho }(t')]}\tilde{O}_t],
\end{equation}

Exploiting the cyclic property of traces:

\begin{equation}\label{expect}
\exx{O}_t -\exx{O}_0 =-i\int_0^t{dt' \text{Tr}[ \tilde{\rho }(t')[\tilde{O}_t,\tilde{H}_I(t')]}],
\end{equation}

Substituting $O$ for the relevant operator eg. $\tilde{\rho}_{00}(t)$, evaluating the commutator, and tracing over the qubit degrees of freedom yields:

\begin{eqnarray}\label{eqn:efmo}
&\exx{\rho_{00}(t)} -\exx{\rho_{00}(0)}=  -i \frac{\Delta}{2} \int_0^t dt'(\exx{\rho_{10}(t')} -\exx{\rho_{01}(t')}), \\
&\exx{\rho_{11}(t)}-\exx{\rho_{11}(0)}=  \phantom{-}i \frac{\Delta}{2} \int_0^t dt'(\exx{\rho_{10}(t')} -\exx{\rho_{01}(t')}),\\
&\exx{\rho_{01}(t)}-\exx{\rho_{01}(0)}=  -i \frac{\Delta}{2} \int_0^t dt' e^{ i\epsilon (t-t')}(\exx{\rho_{00}(t')D_{t} D^{\dagger}_{t'}} -\exx{\rho_{11}(t')D^{\dagger}_{t'}D_{t}}),\\
&\exx{\rho_{10}(t)}-\exx{\rho_{10}(0)}=  \phantom{-}i \frac{\Delta}{2} \int_0^t dt' e^{-i \epsilon (t-t')}(\exx{\rho_{00}(t')D_{t'}D^{\dagger}_{t}} -\exx{\rho_{11}(t')D^{\dagger}_{t}D_{t'}}),
\end{eqnarray}

At this point we make the Born approximation (assuming the density matrix of system and bath are factorable)
\begin{equation}
\exx{\rho_{00}(t') D_t(\alpha) D_{t'}(\alpha)}_{t'} \approx \exx{\rho_{00}(t')}\exx{D_t(\alpha) D_{t'}.(\alpha)}
\label{eqn:born_approx}
\end{equation}
The bosonic correlation function is defined as $C(t-t')$: 
\begin{equation} \label{eqn:bcf}
C(t -t') = \exx{D_{t}(\alpha) D_{t'}^{\dagger}(\alpha)} =   \text{Tr}_B [ \rho_B D_t(\alpha) D_{t'}^{\dagger}(\alpha)]. 
\end{equation}
Where the subscript $B$ represents the bosonic degrees of freedom. We substitute this into (\ref{eqn:efmo}) and by assuming there is no initial coherence in the system we obtain (\ref{eqn:psol1} - \ref{eqn:psol4}).

\subsection{ \label{app:B} Correlation Function}

The bosonic correlation function (\ref{eqn:bosoncf}) for an oscillator with a single mode in a thermal state is defined as:
\begin{equation}
C(t -t') = \text{Tr}_B [ \rho_B D_t( \alpha) D_{t'}^{\dagger}(\alpha)],  \qquad \rho_B = \frac{\exp(- \beta \omega \no)}{\text{Tr}_B [\exp(- \beta \omega \no)} = \frac{1}{Z} \exp(- \beta \omega \no). \label{eqn:bcf2}
\end{equation}
This can be evaluated in different ways, one of which is presented below. Starting from the time dependence of the displacement operator in the interaction picture:
\begin{equation}
D_t(\xi) = e^{i H_0 t} D(\xi) e^{-i H_0 t} = e^{i \omega \no t} D(\xi) e^{-i \omega \no t}, \label{eqn:doip}
\end{equation}
or, alternatively, through the time dependence of creation and annihilation operators:
\begin{equation}
D_t(\xi)  = e^{\xi \cre e^{i \omega t} - \xi^* \an e^{-i \omega t}} = D(\xi e^{i \omega t}). \label{eqn:dotd}
\end{equation}

In order to perform the trace $\text{Tr}_B$ in the number state basis, we need to know the action of $e^{\xi \no}$ and $D(\xi)$ on a number state $\ket{n}$. The first simply evaluates to $e^{\xi n}$ and the latter gives the so-called displaced number state $\ket{\xi, n}$.

The displaced number state can be expanded in the number state basis
\begin{equation}
\ket{\xi, n} = \sum_{m=0}^{\infty} C_{nm} \ket{m}, \qquad C_{nm} = \bra{m} D(\xi) \ket{n}, \label{eqn:dnstate}
\end{equation}
with (see, e.g., Oliviera et al \cite{oliveira05} or M. Crisp \cite{crisp92})
\begin{equation}
C_{nm} = \sqrt{\frac{n!}{m!}} e^{-\frac{1}{2} \abs{\xi}^2} \xi^{m -n} L_n^{m-n}(\abs{\xi}^2), 
\label{eqn:dncoeffs}
\end{equation}
where $L_n^{m-n}(\abs{\xi}^2)$ is an associated Laguerre polynomial. This is only valid for $m > n$, but for $m < n$ the displacement operator, or rather its hermitian conjugate, can be made to act on $\bra{m}$ instead of on $\ket{n}$. 

\subsubsection{Single series and analytical result}

We use Eq. (\ref{eqn:dotd}) for the displacement operator and the property $D(x) D(y) = \exp[(xy^* - yx^*)/2] D(x + y)$ to evaluate Eq. (\ref{eqn:bcf2}). This leads to a series of the following form
\begin{equation}
C(t-t') = \frac{1}{Z} e^{-\abs{\alpha}^2 [1 - e^{-i \omega (t -t')}]} \sum_{n = 0}^{\infty} e^{-\beta \omega n} L_n[2 \abs{\alpha}^2 (1 - \cos[\omega (t-t')])]. \label{eqn:bcf-single-series}
\end{equation}
By virtue of the property $ \sum_{n = 0}^{\infty} L_n(y) z^n  = (1-z)^{-1} \exp[yz / (z-1)]$ and with $N = (e^{\beta \omega} -1)^{-1}$ and $Z = (1 - e^{-\beta \omega})^{-1}$ we finally arrive at the analytical result:
\begin{equation}
C(t -t') = \exp[-i\abs{\alpha}^2\sin[\omega (t-t')]] \exp[-2\abs{\alpha}^2(1-\cos[\omega (t-t')])(N+1/2)]. \label{eqn:bcf-analytical}
\end{equation}
Note that this expression agrees with Mahan's result for a single mode (Ref \cite{mahan00}, section 4.3)
$C(t) = e^{-\abs{\alpha}^2 \left( (1-\cos \omega t) \coth \left( \frac{\beta \omega}{2} \right) + i \sin \omega t \right)}$.
Mahan derives this in a similar fashion but without using Eqs. (\ref{eqn:dnstate}, \ref{eqn:dncoeffs}). Instead, he uses the `Feynman disentanglement of operators' to arrive at an equivalent infinite series of Laguerre polynomials.

\end{document}